\documentclass[preprint,aps,nofootinbib]{revtex4}
\usepackage{graphicx}
\usepackage{amsmath}
\usepackage{amsfonts}
\usepackage{amssymb}
\usepackage{color}%
\providecommand{\U}[1]{\protect\rule{.1in}{.1in}}

\newcommand{\f}{\begin{equation}}
\newcommand{\ff}{\end{equation}}
\newcommand{\fa}{\begin{eqnarray}}
\newcommand{\ffa}{\end{eqnarray}}

\begin{document}
\title{Thermal fluctuations in viscous cosmology}
\author{Wei-Jia Li$^{1}$}
\email{li831415@163.com}
\author{Yi Ling$^{1}$}
\email{yling@ncu.edu.cn}
\author{Jian-Pin Wu$^{2}$}
\email{jianpinwu@yahoo.com.cn}
\author{Xiao-Mei Kuang$^{1}$}
\email{xmeikuang@yahoo.cn} \affiliation{ $^1$Center for Relativistic
Astrophysics and High Energy Physics, Department of Physics,
Nanchang University, 330031, China\\$^2$Department of Physics,
Beijing Normal University, Beijing 100875, China}

\begin{abstract}

In this paper we investigate the power spectrum of thermal
fluctuations in very early stage of viscous cosmology. When the
state parameter as well as the viscous coefficient of a barotropic
fluid is properly chosen, a scale invariant spectrum with large
non-Gaussianity can be obtained. In contrast to the results
previously obtained in string gas cosmology and holographic
cosmology, we find the non-Gaussianity in this context can be
$k$-independent such that it is not suppressed at large scale, which
is expected to be testified in future observation.
\end{abstract}
\maketitle

\section{Introduction}
Recent observed Cosmic Microwave Background (CMB)
anisotropy\cite{CMBobserve,WMAP,WMAPcon2} and the large scale
structure of the
universe\cite{Mukhanov81,Guth82,Hawking82,Starobinsky82,Bardeen83}
can be viewed as the results of the primordial density
perturbations, which is characterized by a nearly scale invariant
and Gaussian power spectrum\cite{WMAP5}. Though at present
completely understanding the origin of the primordial density
perturbations is still an open question, one traditional point of
view is thinking of quantum fluctuations in vacuum as the seed of
such classical perturbations. Nevertheless, there is an
alternative conjecture arguing that it may be due to thermal
fluctuations of matter sources during the inflationary
stage\cite{peeb,steph,rob,mag-pog}. Unfortunately, thermal
fluctuations in standard inflation models always generate a power
spectrum with spectral index $n_s=4$
\cite{thermalmilne,thermalloop}. Recently progress has been made
to overcome this difficulty. People show that once some new
physics is introduced, for instance in the context of the
noncommutative inflation, holographic cosmology and loop
cosmology, the scale invariant spectrum can be implemented in the
thermal scenario as
well\cite{steph,thermalSCG,thermalmilne,thermalloop,thermalholography1,non-commutative1,non-commutative2,ddrlqc}.

Recently the thermal origin of the primordial density perturbation
has received more attention since the non-Gaussianity of CMB has
been further disclosed in the latest observation data. It indicates
that at 95\% confidence level, the primordial non-Gaussianity
parameters for the local and equilateral models are in the region
$-9<f_{\rm NL}^{local}<111$ and $-151<f_{\rm NL}^{equil}<253$,
respectively \cite{WMAP5}. If this result is confirmed by future
experiments such as the Planck satellite, then it will be a great
challenge to many slow-roll inflation models since the
Non-Gaussianity has to be greatly suppressed at $|f_{NL}|<1$ in most
of these models\cite{NG1,NG2}. Contrasting to fluctuations
originated from vacuum, the fluctuations in thermal scenario are
always not strictly Gaussian. This may provide an alternative way to
seek the observable non-Gaussianity. Therefore, the thermal
non-Gaussianity has been widely discussed in recent
literatures\cite{thermalinflation,thermalSGC2,Ling,Ling2,LM,Cai,Gao,Byrnes,Lin,Battefeld,Chen,Kumar}.

In this paper we intend to investigate the power spectrum of thermal
fluctuations in the very early stage of viscous cosmology, where the
matter source is a viscous barotropic fluid (the state parameter
$w=\frac{p}{\rho}\in[-1,1]$ is a constant). Since the viscous effect
has an impact on the perturbation modes after they cross the thermal
horizon, it is expected that the power spectrum should be different
from the previous results even without other new mechanism or extra
structure introduced. As a matter of fact, we find that provided the
state parameter as well as the viscous coefficient of a barotropic
fluid is properly chosen, a scale invariant spectrum with large
Non-Gaussianity can be obtained in this framework indeed.

The outline of our paper is the following. In next section we
present a brief review on the thermal fluctuations in the FRW
universe filled with a perfect fluid and demonstrate the
difficulty of gaining a scale-invariant spectrum in this model.
Then in section \textrm{III} we show that this difficulty can be
overcome by introducing a viscous fluid and properly choosing its
viscous coefficient. Its non-Gaussianity is investigated in
section \textrm{IV}. Finally the comparison with other
cosmological models with thermal fluctuations is given in the
section of Conclusion and Discussion.

\section{A review on thermal fluctuations in the FRW universe with a perfect fluid}
In this section, we briefly review the thermal fluctuations in
very early universe and derive its power spectrum based on
thermodynamical consideration. First of all, we consider the FRW
universe filled with a perfect fluid with state parameter $w$. For
a perfect fluid, the thermal system can reach equilibrium state
through interactions at sub-Horizon scales. In the thermal
scenario of cosmological fluctuations, this process is called
thermalization. Since during this process both of the extensive
internal energy $U$ and entropy $S$ are well defined for a global
equilibrium state, we have the following form of the first law of
thermodynamics
\begin{equation}\label{d}
dU=TdS-pdV.
\end{equation}
For barotropic fluid, one finds that the relation between the energy
density and the temperature is uniquely fixed by this
thermodynamical law\cite{thermalloop,thermalmilne}
\begin{equation}\label{h}
\rho=AT^{m},
\end{equation}
where $m=1+\frac{1}{w}$ and $A$ is an integral constant. For
radiation with $w=1/3$, this equation is nothing but the famous
Stefan-Boltzmann law. On the other hand, the partition function of
the thermal fluid is defined as
\begin{equation}\label{i}
Z={\sum_r} e^{-\beta E_r} \ ,
\end{equation}
where $\beta = T^{-1}$. The internal energy $U$ inside a volume
$V$ is given by
\begin{equation}\label{j}
U={\langle E\rangle}={{\sum _r} E_r e^{-\beta E_r} \over {\sum_r}
e^{-\beta E_r}}=-{d\log Z\over d\beta}.
\end{equation}
Therefore, the two-point correlation function for the density
fluctuation $\delta\rho\equiv\rho-\langle\rho\rangle$ can be
obtained as
\begin{equation}\label{k}
\langle\delta\rho^2\rangle=\frac{\langle\delta
E^2\rangle}{V^2}=\frac{{\langle E^2\rangle}-{\langle
E\rangle}^2}{V^2}=\frac{1}{V^2}{d^2\log Z\over d\beta^2}=
-\frac{1}{V^2}{dU \over d\beta}= \frac{T^2C_V}{R^6},
\end{equation}
where $C_V=(\frac{\partial U}{\partial
T})_{V}=V\frac{d\rho}{dT}\equiv V \rho'$ is the specific heat and
$R\sim V^{\frac{1}{3}}$ is the size of the thermal horizon. Thermal
fluctuations generated from the matter inside $R$ can be described
by the thermodynamics above. But when thermal modes are pushed
outside the horizon, they are frozen and become non-thermal governed
by the theory of perturbations. Next we calculate the power spectrum
of perturbations. Following most of literatures we identify the
thermal horizon $R$ with the Hubble horizon
$H^{-1}$\cite{thermalSCG,thermalmilne,thermalloop,thermalholography1,Ling,Ling2,Cai},
which means our calculation of spectrum is always taken at the
Hubble scale.

If perturbations are deeply in the horizon, the 0-0 component of
the perturbative Einstein equation will reduce to the Poisson
equation which relates the curvature fluctuations $\Phi_k$ and the
density perturbations $\delta \rho_k$ as\cite{Liddle}
\begin{equation}\label{l}
k^2\Phi_k=4\pi G a^2\delta\rho_k,
\end{equation}
where $\delta\rho_{k}=k^{-{3\over2}}\delta\rho$. Thus the power
spectrum can be obtained by combining Eq.(\ref{k}) and (\ref{l})
\begin{equation}\label{o}
{\cal P}_\Phi(k)\equiv\frac{k^3}{2\pi}\langle\Phi_{k}^{2}\rangle
\sim {a\over k}T^2 \rho',
\end{equation}
where we have used the condition $R=H^{-1}=\frac{a}{k}$. When the
$\Phi$ modes leave the horizon, their amplitudes get fixed at
whatever thermal amplitudes they have at crossing $k=aH$. For
simplicity, we consider the spatially flat universe. Then using the
Friedmann equation $H^2\propto \rho$, we have
\begin{equation}\label{p}
{\cal P}_\Phi(k)\sim {\left[T^2\rho'\over
\sqrt{\rho}\right]}_{k=aH}.
\end{equation}
For a constant $w$, substituting the thermodynamical relation
(\ref{h}) into above equation leads to
\begin{equation}\label{q}
{d\ln {\cal P}_\Phi\over d\ln T}=1+{m\over 2}.
\end{equation}
Furthermore, from the conservation equation of the fluid one has
$\rho\propto a^{-3(1+w)}$ such that
\begin{equation}\label{r}
a\propto T^{-m\over 3(1+w)}.
\end{equation}
Thus we have
\begin{equation}\label{s}
{d\ln k\over d\ln T}={m(1+3w)\over 6(1+w)},
\end{equation}
where we have used $k=aH\propto a\sqrt{\rho}$. Using (\ref{q}) and
(\ref{s}), it is straightforward to calculate the spectral index as
\begin{equation}\label{s1}
n_S-1={d\ln {\cal P}_\Phi\over d\ln k}= {d\ln {\cal P}_\Phi\over
d\ln T}{d\ln T\over d\ln k}=3\frac{2+m}{m}\frac{w+1}{3w+1}=3.
\end{equation}
In this equation the relation $m=1+\frac{1}{w}$ has been applied.
Therefore the spectrum is always blue and independent of the value
of $w$. Of course this ``\textit{no-go result}''\cite{thermalloop}
is not consistent with the current experiments, in which $n_s$ is
restricted at $n_s=0.960^{+0.014}_{-0.013}$ \cite{WMAP5}. Then, to
obtain a scale invariant spectrum in thermal scenario, one need
introduce new physics to either relax some constraints due to
thermodynamics, for instance as presented
in\cite{thermalmilne,thermalholography1}, or modify the standard
cosmological equations, as presented in \cite{thermalloop}.
Different from above considerations, in next section we would like
to argue that the \textit{no-go result} above can also be avoided
if the matter source is a viscous fluid rather than a perfect one.

\section{realization of scale-invariant fluctuations in viscous cosmology}
Viscous cosmology has been widely applied to investigate the
structure and the evolution of the universe
\cite{Barrow,Davies,Kremer,Brevik,Cataldo,Nojiri,Feng,Kuang,Gron,Maartens1,Maartens2,Campo,Brevik3,Barrow2,Tawfik}(For
recent review we refer to \cite{Zimdahl}). In this context the
energy-momentum tensor of the fluid is given by
\begin{equation}\label{a1}
T_{\mu\nu}=(\rho+p-3\xi H)u_\mu u_\nu+(p-3\xi H) g_{\mu\nu},
\end{equation}
where the bulk viscosity coefficient $\xi=\xi(\rho)$ is usually a
function of the energy density $\rho$ of the fluid. Moreover, as
pointed out in \cite{Weinberg}, $\xi$ should be positive if the
second law of thermodynamics is respected. In $FRW$ universe the
conservation equation $\triangledown^\mu T_{\mu\nu}=0$ becomes
\begin{equation}\label{b}
\dot{\rho}+3H(\rho+p-3\xi H)=0.
\end{equation}
In this paper, we choose a special kind of viscous fluids with a
bulk viscosity coefficient $\xi(\rho)\propto\rho^{\frac{1}{2}}$,
which has already been investigated in some literatures
\cite{Brevik,Paolis,Kuang}. In this case, $\frac{3\xi
H}{\rho}\equiv\alpha $ is a positive constant which is relevant to
the viscosity of the fluid. Then given a constant $w$, we have a
relation between the energy density $\rho$ and the scale factor from
the equation above
\begin{equation}\label{c}
\rho \propto a^{-3(1+w-\alpha)}.
\end{equation}
Obviously, the evolution of the universe depends not only on the
state parameter $w$ but also on $\alpha$.

Now we discuss the thermalization of the matter source. Usually,
the thermal scenario of cosmological fluctuations is established
on the thermodynamics of the equilibrium state. Unlike the case
with a perfect fluid, the viscosity brings dissipative effect such
that the global description of the first law of thermodynamics in
Eq.(\ref{d}) no longer stands in general. Nevertheless, each
particle in a viscous fluid satisfies the Gibbs relation in a
local equilibrium state
\begin{equation}\label{c1}
Tds=d\frac{\rho}{n}+pd\frac{1}{n},
\end{equation}
where $s$ is the entropy and $n$ is the particle number in a local
equilibrium element. For the barotropic case, it can be rewritten as
\begin{equation}\label{ADD1}
ds=\frac{1}{Tn}d\rho-\frac{\rho(1+w)}{Tn^2}dn.
\end{equation}
Employing the integrability condition
\begin{equation}\label{ADD2}
\frac{1}{T}\Big(\frac{\partial}{\partial
n}\frac{1}{n}\Big)_\rho=-\frac{(1+w)}{n^2}\Big(\frac{\partial}{\partial
\rho}\frac{\rho}{T}\Big)_n,
\end{equation}
one can still derive the Stefan-Boltzmann law $\rho \propto T^{m}$
with $m=1+\frac{1}{w}$ \cite{Maartens1,Maartens3}. This result
indicates the viscosity does not change the relation between the
energy density and the temperature of the fluid.

In a heuristic manner we may argue that the anisotropy and the
inhomogeneity of our universe totally originate from the
fluctuations in such a local equilibrium element. Consider a quanta
of the fluctuation with a momentum $P_r$. Since the momentum
uncertainty of a quantum particle $\Delta P_r$ is of order of
$P_r$\cite{Adler,Lingc,Han}, from the uncertainty principle we have
\begin{equation}\label{ADD3}
P_r=\hbar k_{rph}\sim \Delta P_r\sim\frac{\hbar}{\Delta x_r},
\end{equation}
where $k_{rph}=\frac{k_r}{a}$ is the physical wavenumber. Then we
have $\Delta x_r\sim\frac{a}{k_r}$. Using the crossing condition
$k_r=a_rH_r$, we derive its position uncertainty $\Delta x_r\sim
H_r^{-1}$. Since all of the fluctuations happen in the local
equilibrium element, the position uncertainty should not be larger
than the scale of the element $\Delta x_r\leq R$. We find
\begin{equation}\label{ADD4}
H^{-1}\sim \Delta x\leq R,
\end{equation}
when each $k$ crosses the horizon. In other words, this local
equilibrium element within a radius of the Hubble scale can be
viewed as the cradle of the cosmological perturbations. Moreover,
we point out that both the background equation $H^2\propto\rho$
and the Poisson equation (\ref{l}) will not change due to the
viscosity effects and this can be understood from the fact that
the viscous term does not appear in the 0-0 component of the
energy-momentum tensor, as shown in Eq.(\ref{a1}). Therefore the
correlation functions can be similarly calculated as in previous
section. Inserting the relation in Eq.(\ref{h}) into (\ref{c}), we
find the variation of the scale factor with the temperature is
modified as
\begin{equation}\label{c2}
a\propto T^{-m\over 3(1+w-\alpha)}.
\end{equation}
Repeating the calculation in previous section, we derive
 \begin{equation}\label{s2}
{d\ln k\over d\ln T}={m(1+3w-3\alpha)\over 6(1+w-\alpha)}.
\end{equation}
As a result, the spectral index for a viscous fluid can be
obtained as
\begin{equation}\label{s3}
n_S-1={d\ln {\cal P}_\Phi\over d\ln k}=
3\frac{2+m}{m}\frac{w-\alpha+1}{3w-3\alpha+1}=3\frac{3w+1}{w+1}\frac{w-\alpha+1}{3w-3\alpha+1}.
\end{equation}
We find that the effect of the viscosity does imprint on the power
spectrum of the cosmological perturbations. Especially, if we
choose $w\approx \alpha-1$ or $w\approx -\frac{1}{3}$ with any
non-vanishing $\alpha$, a nearly scale-invariant power spectrum of
the perturbations can be implemented. The case with $w\approx
\alpha-1$ is nothing but the standard inflation, because
$\dot{H}=-4\pi G\rho(1+w-\alpha)\approx0$ implies a nearly de
Sitter expansion. When $w\approx -\frac{1}{3}$, since $\alpha>0$
as required by the second law of thermodynamics, we have
$\frac{\ddot{a}}{a}=-4\pi G \rho(1+3w-3\alpha)>0$. In this case,
the universe undergoes an accelerating expansion as well.
Therefore, in contrast to models with a perfect fluid, the thermal
scenario in viscous cosmology may provide an alternative
interpretation on the scale invariance of the power spectrum.
However, to be well consistent with the current observation
constraints, we need further study the non-Gaussianity in this
thermal scenario, and this is what we intend to do in next
section.

\section{Thermal non-Gaussianity in viscous cosmology}

Another important quantity which is expected to be measured
precisely in coming experiments is the non-Gaussianity of
\textit{CMB}. The amount of the non-Gaussianity is usually
estimated by the quantity $f_{NL}$ which is defined by the
three-point correlation function for the curvature perturbations
$\langle\zeta_{k_1}\zeta_{k_2}\zeta_{k_3}\rangle$. Generally, two
cases are considered. one is the local non-Gaussianity labeled by
$f_{NL}^{local}$ with $k_1\ll k_2\approx k_3$, and the other is
the equilateral non-Gaussianity denoted by $f_{NL}^{equil}$ with
$k_1\approx k_2\approx k_3$.  More detailed discussions on these
quantities can be found in\cite{shape,shapeli}. Recent WMAP-5 data
shows the possibility that there exist large non-Gaussianities of
cosmological perturbations\cite{WMAP5}. If this is true, thermal
scenario may give a more consistent result. In this section, we
will calculate the equilateral non-Gaussianity of thermal
fluctuations in the context of viscous cosmology.

The non-Gaussianity is quantificationally described by the
non-Gaussianity estimator
\begin{equation}\label{fdefine}
  f_{NL}^{equil} = \frac{5}{18}k^{-\frac{3}{2}}\frac{\langle {\zeta}_{k}^{3}
  \rangle}{\langle{\zeta}_{k}^{2}\rangle
  \langle{\zeta}_{k}^{2}\rangle}~,
\end{equation}
where $\zeta_k$ is the \textsl{curvature perturbation on slices of
uniform energy density}, while $\langle {\zeta}_{k}^{2}\rangle$ and
$\langle {\zeta}_{k}^{3}\rangle$ are two-point and three-point
correlation functions, respectively. Since no entropic perturbations
occur in single component case\cite{Riotto}, the cosmological
perturbations are adiabatic. Then $\zeta$ can be given as
\begin{equation}\label{z}
\zeta=\Phi+H\frac{\delta\rho}{\dot{\rho}}.
\end{equation}
It remains constant on super-horizon scales in the adiabatic case
even if $w$ is changing. Considering the fact that
$\frac{\delta\rho}{\rho}$ is always equal to $-2\Phi$ on
super-horizon scales\cite{Mukhanov} and using (\ref{b}), we can
rewrite Eq.(\ref{z}) as
\begin{equation}\label{Phizeta2}
\zeta=\frac{5+3w-3\alpha}{3+3w-3\alpha}\Phi.
\end{equation}
Since both $w$ and $\alpha$ are constants, $\Phi$ is also a
constant. Therefore, we can obtain $f_{NL}^{equil}$ by evaluating
$\langle\Phi_{k}^{2}\rangle$ and $\langle\Phi_{k}^{3}\rangle$.
From Eq.(\ref{h}), we have the two-point correlation
\begin{equation}\label{rho2}
  \langle\delta\rho^2\rangle=\frac{\langle\delta
  E^2\rangle}{V^{2}}=-\frac{1}{V^2}{dU \over d\beta}=\frac{mAT^{m+1}}{V},
\end{equation}
and the three-point correlation
\begin{equation}\label{rho3}
  \langle\delta\rho^3\rangle=\frac{\langle\delta
  E^3\rangle}{V^{3}}=\frac{1}{V^3}{d^2U \over d\beta^2}=\frac{m(m+1)AT^{m+2}}{V^{2}}.
\end{equation}
Applying the horizon crossing condition $k=a/R=aH$ and the Poisson
equation (\ref{l}), the two-point correlation function for $\Phi_k$
is given as
\begin{equation}\label{Phi2}
  \langle\Phi_{k}^{2}\rangle =(4 \pi G)^{2}R^{4}\langle \delta\rho_{k}^{2}\rangle=(4 \pi
  G)^{2}R
  k^{-3}mAT^{m+1},
\end{equation}
while the three-point correlation function is
\begin{equation}\label{Phi3}
  \langle\Phi_{k}^{3}\rangle =(4 \pi G)^{3}R^{6}\langle \delta\rho_{k}^{3}\rangle=
  (4 \pi G)^{3}k^{-\frac{9}{2}}m(m+1)AT^{m+2}.
\end{equation}
Then the non-Gaussianity estimator $f_{NL}^{equil}$ can be
calculated with the use of Eq.(\ref{fdefine}). It turns out
\begin{equation}\label{f1}
f_{NL}^{equil} = \frac{5}{72\pi G AT^m
R^2}\frac{3+3w-3\alpha}{5+3w-3\alpha}\frac{(m+1)}{m}
  =\frac{5}{72\pi G\rho
R^2}\frac{3+3w-3\alpha}{5+3w-3\alpha}\frac{2w+1}{w+1}.
\end{equation}
Obviously the amount of non-Gaussianity also depends on both $w$ and
$\alpha$. In the previous section, we have concluded that the state
parameter should be restricted at $w\approx\alpha -1$ or
$w\approx-\frac{1}{3}$ in order to obtain a scale invariant
spectrum.

In the case of $w\approx\alpha-1$ , we get a small non-Gaussianity
\begin{equation}\label{f2}
f_{NL}^{equil}\approx0,
\end{equation}
which is not surprising since it leads to de-Sitter expansion as
in the slow-roll inflation scenario.

In the case of $w=-\frac{1}{3}$, we have
\begin{equation}\label{f3}
f_{NL}^{equil} = \frac{5}{144\pi G\rho
R^2}\frac{2-3\alpha}{4-3\alpha}=\frac{5}{54}\frac{2-3\alpha}{4-3\alpha}.
\end{equation}
As a result, if we choose the value of $\alpha$ close to
$\frac{4}{3}$, a large non-Gaussianity can be obtained.
Furthermore, we find the non-Gassianity is scale-invariant as
well. That is to say, it is not suppressed at large scale, which
is in contrast to the previous results in string gas cosmology and
holographic cosmology where the non-Gaussianity is $k$
dependent\cite{thermalSGC2,Ling}. Finally, it is worthwhile to
point out that in this case we have $\dot{H}>0$, which implies a
phantom like universe.

\section{conclusion and discussion}
In this paper, we have investigated the power spectrum of thermal
fluctuations in very early stage of viscous cosmology with a special
viscous coefficient. Firstly, with the conjecture that the origin of
the structure of our universe is completely thermal, we have
demonstrated that a nearly scale invariant spectrum can be achieved
when the state parameter and the viscous coefficient are properly
selected. Different from previous literatures, we have only
introduced the viscosity of the fluid, without modifying the
Einstein equation or the thermodynamical relations of matter
sources. Moreover, the results of thermal scenario under the
condition of $w\approx\alpha-1$ is closely analogous to those
obtained in standard inflation scenario.

Secondly, we analyze the non-Gaussianity of the thermal fluctuations
in this scenario. It is shown that there will be a suppressed
non-Gaussianity in the case $w\approx\alpha-1$. It is just analogous
to the usual inflationary phase where $w\approx-1$ is required.
However, if we choose $w\approx-\frac{1}{3}$, a large and
$k$-independent non-Gaussianity can be achieved by tuning $\alpha$
near $\frac{4}{3}$ which implies a phantom like universe. It is of
interesting to notice that a phantom phase is also required in
holographic cosmology and the near Milne cosmology when a large
non-Gaussianity can be obtained.

In conclusion, due to the viscosity effects of non-perfect fluids,
the \textit{no-go result} of the thermal fluctuations can be
avoided and a large non-Gaussianity can also be achieved.
Therefore it is desirable to investigate furthermore the influence
of viscosity effects on the structure formation and anisotropy of
the \textit{CMB}.

\section*{Acknowledgement}
W. J. Li, Y. Ling and X. M. Kuang is partly supported by
NSFC(Nos.10663001,10875057), JiangXi SF(Nos. 0612036, 0612038), Fok
Ying Tung Education Foundation(No. 111008) and the key project of
Chinese Ministry of Education(No.208072). J. P. Wu is partly
supported by NSFC(No.10975017). We also acknowledge the support by
the Program for Innovative Research Team of Nanchang University and
Jiangxi Young Scientists(JingGang Star) Program.

\end{document}